\documentclass[prl,superscriptaddress,showpacs,twocolumn,floatfix]{revtex4-1}

\usepackage{graphicx}
\usepackage{dcolumn}
\usepackage{bm}
\usepackage{amsmath} 

\begin{document}
\title{The single-particle unit for alpha decay}
\author{Chong Qi}\email{chongq@kth.se}
\affiliation{Royal Institute of Technology at Albanova,
            SE-10691 Stockholm, Sweden }
\author{Roberto J. Liotta}
\affiliation{Royal Institute of Technology at Albanova,
	SE-10691 Stockholm, Sweden }
\author{Ramon Wyss}
\affiliation{Royal Institute of Technology at Albanova,
	SE-10691 Stockholm, Sweden }
\date{\today}

\begin{abstract}

A salient feature of quantum mechanics is the inherent property of collective quantum motion, 
when apparent independent quasiparticles move in highly correlated trajectories, 
resulting in strongly enhanced transition probabilities. 
To assess the extend of a collective quantity requires an appropriate 
definition of the uncorrelated 
average motion, often expressed by single particle units.
A well known example in nuclear physics is the Weisskopf unit for electromagnetic 
transitions which reveals different aspects of collective motion. 
In this paper we define the corresponding single particle unit for alpha decay 
following Weisskopf's derivations. 
We evaluate the alpha decay amplitude as induced by four 
uncorrelated/non-interacting protons and 
neutrons and compare it with the one extracted from observed decay rates.
Our definition elucidates the collectivity in
alpha decay and facilitates an unified description of all alpha decay 
processes along the nuclear chart. 
Our formalism is also applicable to other particle decay processes.
 
\end{abstract}


\maketitle
Alpha decay has been one of the most rewarding subjects in physics
since Gamow was the first to apply the probabilistic interpretation of quantum 
mechanics to describe the penetrability of the 
Coulomb barrier by the $\alpha$-particle~\cite{gam28}. The subsequent developments upon radioactive particle
decay in nuclear physics has been outstanding~\cite{Lovas, clw}. At present, 
$\alpha$-decay is crucial for the identification of unstable nuclei far from 
stability, particularly super heavy and proton rich nuclei~\cite{VDA}.
Yet there are unsolved fundamental problems even today: A microscopic description of
the clustering of the four
nucleons which eventually constitute the $\alpha$-particle
as provided by the nuclear configuration interaction shell model has not been
able to account for the decay widths, even when
high-lying configurations as well as 
the proper treatment of the nuclear continuum is incorporated into the calculations.

The understanding and quantification of collective motion in atomic nuclei
have a long history. Enhanced decay probabilities in electromagnetic transitions
are used to classify different excitation modes such as vibrations and rotations. 
These classifications of collectivity can be made through a reliable basic quantity, namely the 
single-particle Weisskopf unit (W.u.) \cite{wei}. Such a common reference 
enables one to differentiate between decays that are non collective and those 
that involve the coherent motion of many nucleons. 
Although called "unit", the W.u. has not an universal value, since it 
depends upon the mass of the nucleus in question as well as upon the character 
of the transition ($E\lambda$ or $M\lambda$). 

Analogous to the W.u. for electromagnetic decay, we define in this letter 
an equivalent unit for  $\alpha$ decay, the particle decay unit, p.d.u. This will relate the measured 
probability of  $\alpha$ decay to an averaged single configuration in the 
description of the mother nucleus. We hope that this common reference will 
clarify the role played by nuclear collectivity in  $\alpha$ decay. 
Our definition enables the appropriate comparison of all hitherto observed 
$L=0$  $\alpha$ decay on the same footing, avoiding the multitudes of 
effective quantities found at present in the 
literature~\cite{Ras59,Hil19,Cub20}. In addition, the formalism presented 
in this paper will enable one to quantify the role 
played by $\alpha$ clustering in heavy nuclei.

Below we present in detail the formalism. We start with the Thomas expression
for the  $\alpha$ decay width~\cite{Thom},
\begin{equation}\label{Thomas}
\Gamma_c (R) = \hbar/T = \frac{\hbar^2 k}{\mu}
\frac{R^2 |{ F}_c(R)|^2}{|H_l^+(\chi,\rho)|^2}
\end{equation}
which often is written as
\begin{equation}\label{widef}
\Gamma_c (R) =  \frac{\hbar^2 R}{\mu}  |{F}_c(R)|^2 P_c(R)
\end{equation}
where 
$
P_c(R)={kR}/{|H_l^+(\chi,\rho)|^2}
$
is the penetrability of the already formed  $\alpha$ particle through the Coulomb and centrifugal 
fields starting at the point $R$, which  is the 
distance between the mass centres of the daughter nucleus and  $\alpha$ cluster.
In these equations $c$ labels the decaying channel, $k$ is the linear momentum carried
by the $\alpha$-particle, $\mu$ is the reduced mass,  
 $H^+_l$ is the Coulomb-Hankel function describing the two-body system in
 the outgoing channel. Its arguments are $\rho=\mu\nu R/\hbar$ and 
$\chi = 2Z_cZ_de^2/\hbar\nu$.  $Z_c$ and $Z_d$ are
the charge numbers of the cluster and daughter nucleus, respectively. The function
 ${F}_c(R)$ is the  $\alpha$ formation amplitude, i.e., 
the mother wave function describing the motion of the $\alpha$ cluster in the 
field induced by the daughter nucleus at the point $R$. 
It is important to stress the difference between this exact treatment and the
effective treatments mostly used in the literature. In 
Eq. (\ref{Thomas}) the evaluation of the formation amplitude is assumed to be
performed within a microscopic framework~\cite{TonAr,Mang}. At the point $R$ in 
Eq. (\ref{Thomas}) the $\alpha$-particle is 
already formed and only
the Coulomb and centrifugal interactions are relevant.

In  microscopic treatments of the formation amplitude the 
height of the Coulomb barrier does not play any direct role since the motion 
of the four nucleons which eventually constitute the  $\alpha$ particle is 
determined by all nucleon-nucleon interactions inside the mother nucleus. 
The greatest challenge facing microscopic treatments is to describe the 
clustering of the four nucleons at $R$. Once this is solved one evaluates the 
probability that the cluster escapes the mother nucleus according to Eq.~(\ref{widef}). To achieve the microscopic description of the $\alpha$ 
clusterization and the subsequent motion of the cluster at the surface $R$ is a difficult undertaking \cite{Lovas,Mang,TonAr,Delion92}. This explains 
why effective treatments are common in the literature, where the  $\alpha$ particle is assumed to exist inside the 
mother nucleus. The decay is described as the penetration of a preformed  
$\alpha$ particle through the Coulomb barrier. 

In our formalism the formation amplitude is determined following the microscopic treatment \cite{clw}, 
i.e.,     
\begin{equation}\label{foram}
{F}_c(R)=\int d\hat{R} d\xi_d d\xi_{\alpha}
[\Psi_d(\xi_d)\phi_{\alpha}(\xi_{\alpha})Y_l(\hat{R})]^*
\Psi_m,
\end{equation}
where $\xi_d$ and $\xi_{\alpha}$ are the internal degrees of freedom
determining the dynamics of the daughter nucleus and the $\alpha$-particle. 
The wave functions $\Psi_d(\xi_d)$ and
$\Psi_m(\xi_d,\xi_{\alpha},\hat{R})$ correspond to the daughter and
mother nuclei respectively. The $\alpha$-particle wave function has the form of a
$n=l=0$ harmonic oscillator eigenstate in the neutron-neutron relative 
distances $r_{nn}$, as well as in the proton-proton distance $r_{pp}$ and 
in the distance $r_{np}$ between the mass centres of the $nn$ and $pp$ pairs~\cite{clw}, 
\begin{equation}
\phi_{\alpha}(\xi_{\alpha})=\sqrt{\frac{1}{8}}(\frac{\nu_\alpha}{\pi})^{9/4}
exp[-\nu_\alpha(r_{nn}^2+r_{pp}^2+2r_{pn}^2)/4]S_\alpha
\end{equation}
where $S_\alpha$ is the $\alpha$-spinor corresponding to the lowest
harmonic oscillator wave function. The total angular momenta are $L=S=0$.
The quantity  $\nu_\alpha=0.574 fm^{-2}$ is the $\alpha$-particle harmonic 
oscillator parameter \cite{lev}. 

We consider decays involving uncorrelated states of even-even nuclei. We will 
focus our treatment on ground-state to ground-state transitions, implying that
 $l$=0 and $Y_{l=0}(\hat{R})=1/\sqrt{4\pi}$. Uncorrelated decay means that
the mother nucleus consists of the daughter 
nucleus times a pure configuration of a pair coupled to
zero angular momentum times a similar proton pair, i. e.
\begin{equation}
\label{mwf}
\Psi_m(\xi_d,\xi_{\alpha},\hat{R})=(\varphi_\nu(\mathbf{r_1})\varphi_\nu(\mathbf{r_2}))_{00}
(\varphi_\pi(\mathbf{r_3})\varphi_\pi(\mathbf{r_4}))_{00}\Psi_d(\xi_d)
\end{equation}
Writing the single-particle wave functions $\varphi_i(\mathbf{r})$  in
their radial, angular and spin components, these last two are canceled in the angular and spin integrals in Eq. (\ref{foram}). 
In order to perform 
the radial part of this integral it is convenient to write the mother wave function 
in terms of the relative coordinates $\mathbf{r_{nn}}$,
$\mathbf{r_{pp}}$, $\mathbf{r_{pn}}$ and the centre of mass coordinate $\mathbf{R}$. Since
the Jakobian corresponding to the transformation from absolute to relative coordinates in the
integral (\ref{foram}) is unity one can  write
\begin{equation}
\label{relwf}
\Psi_m(\xi_d,\xi_{\alpha},\hat{R})=\phi(\mathbf{r_{nn}})\phi(\mathbf{r_{pp}})
\phi(\mathbf{r_{pn}})\phi(\hat{R})\Psi_d(\xi_d)
\end{equation}
where $\phi$ are the wave functions in relative coordinates. 
These functions  may diverge at $r=0$ and therefore we use the standard 
function $u(r)=r\phi(r)$. Following the method employed by Weisskopf, we 
assume that the radial single-particle wave function $u(r)$ in Eq. (\ref{mwf}) 
is constant inside the mother nucleus, with radius $R$. As a result, the 
relative and centre of mass radial wave functions inside the mother nucleus 
are constants. Notice that according to our prescription the $nn$, $pp$ and 
$pn$ wave functions vanish outside the nuclear surface, while $\phi (R)$, the 
wave function  corresponding to the motion of the $\alpha$ particle centre of 
mass, is constant inside the nucleus, but outside corresponds to
an outgoing  $\alpha$ particle, as seen below.

The normalization condition provides
\begin{equation}\label{norco}
\int_0^{R}(u({r})/r)^2 r^2dr=R C^2=1
\end{equation} 
where the constant $C$ is the same for the ${pp}$, $nn$, $pn$ and the centre 
of mass wave functions inside the mother nucleus resulting in
$C=1/\sqrt{R}$.

The formation amplitude in Eq. (\ref{foram}) acquires the form,
\begin{eqnarray}\label{formation}
{ F}_c (R)&=&\int d\hat{R}\int r_{nn}^2dr_{nn}r_{pp}^2dr_{pp}r_{pn}^2dr_{pn}
\sqrt{\frac{1}{8}}(\frac{\nu_\alpha}{\pi})^{9/4}\nonumber\\
&&\times e^{-\nu_\alpha(r_{nn}^2+
	r_{pp}^2+2r_{pn}^2)/4}\frac{1}{\sqrt{4\pi}}\frac{C^4}{r_{nn}r_{pp}r_{pn}R}
\nonumber\\
& =&\int r_{nn}dr_{nn}r_{pp}dr_{pp}r_{pn}dr_{pn}
\sqrt{\frac{1}{8}}(\frac{\nu_\alpha}{\pi})^{9/4}
\nonumber\\
&&\times e^{-\nu_\alpha(r_{nn}^2+
	r_{pp}^2+2r_{pn}^2)/4}\frac{\sqrt{4\pi}}{R^3}
\end{eqnarray}
It is straightforward to perform  the radial integrals. Thus for $r_{nn}$ one 
obtains,
\begin{equation}
\int r_{nn}dr_{nn}exp[-\nu_\alpha r_{nn}^2/4]=
\frac{2}{\nu_\alpha}.
\end{equation}
The  remaining integrals can be calculated in the same fashion. We are interested 
in the formation amplitude at the radius $R$ and therefore integrate over the 
angle $\hat R$ (which provides a factor $4\pi$). 
The formation amplitude at the nuclear surface becomes,
\begin{eqnarray}\label{pdu}
{ F}_{\alpha;{\rm pdu}}(R)=\sqrt{\frac{1}{8}}(\frac{\nu_\alpha}{\pi})^{9/4}\sqrt{4\pi} 
\frac{C^4}{R}
\frac{4}{\nu_\alpha^{3}}=\frac{\sqrt{8}\nu_\alpha^{-3/4}\pi^{-7/4}}{R^3}
\end{eqnarray}
which defines the particle decay unit (p.d.u.). It measures the  $\alpha$ decay 
formation amplitude for decays from four uncorrelated single particle states. 
With $R=1.2(A^{1/3}+4^{1/3})$~fm one obtains
\begin{eqnarray}\label{pdu2}
{ F}_{\alpha;{\rm pdu}}= {0.335}/{(A^{1/3}+4^{1/3})^3}~ {\rm fm}^{-3/2}.
\end{eqnarray}

In order to clarify the procedure that we are following here it is important to remember
that the neutrons and protons form the  $\alpha$ particle 
at the nuclear surface because of the interactions among 
them inside the mother nucleus. 
At and inside the nuclear surface the 
$\alpha$ particle wave function has  the constant value $u(\mathbf{r})=C$. 
Outside the nuclear surface, i. e. at $r>R$ (where only the Coulomb and 
centrifugal interactions are relevant), 
the wave function of the outgoing  $\alpha$ particle becomes  
\begin{equation}
\label{outg}
u(r)=r\phi(r)=N[H^+_l(\chi,\rho)]
\end{equation}
where $N$ is the matching constant. The independence of the Thomas expression 
upon the distance $R$ (as pointed out above, $R$ should be beyond the nuclear 
surface) has often been used in 
microscopic calculations of  $\alpha$ decay to probe whether the results are 
reliable \cite{for}.

Following Eq.~(\ref{pdu2}), we extract the $\alpha$ decay formation amplitude 
measured in p.d.u. from the ratio between 
experiment and the corresponding p.d.u.value. 
Similar to the W.u. in electromagnetic decay, values that
exceed the p.d.u. by an order of magnitude
reflect the enhancement in  $\alpha$ decay, pointing towards the collectivity of the process.

\begin{figure}
	\begin{center}
		\includegraphics[width=0.45\textwidth]{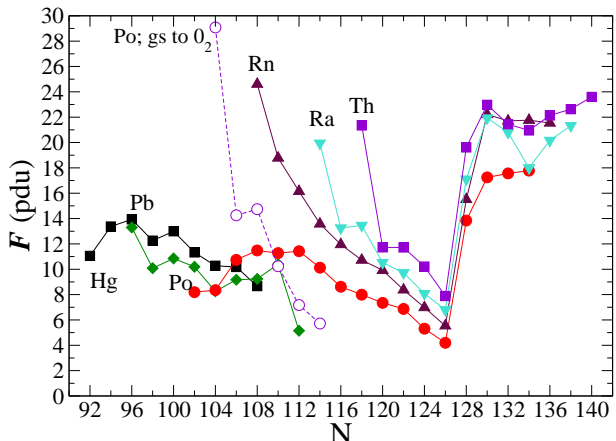}
	\end{center}
	\vspace{-0.8cm}\caption{(color online). $ \alpha$-particle
		formation probabilities in p.d.u. for the decays of the even-even isotopes 
		as a function of the neutron numbers $N$ of the mother nuclei. 
	}
	\label{fap}
\end{figure}

The value of the $\alpha$ formation amplitudes in p.d.u. of known  $\alpha$ 
emitters in the mass $A=180$ region and beyond are depicted in Fig. \ref{fap}. The experimental half-lives are taken 
from Ref. \cite{And13} and references therein.
The figure reveals distinctive features characterizing  $\alpha$ decay. Thus and most conspicuous, the decay rates all 
exceed by far the value of a 
single particle unit. In other words, the  $\alpha$ decay
process in its nature reveals strong collectivity. Hence, it is not surprising that cluster
components are needed in the shell model wave function to account for the experimental decay width~\cite{Lovas, dls}.
Other important feature revealed by the figure is the shell closure at
$N=126$. For heavier isotopes, i. e. above the magic number 126, the
p.d.u. approach a constant value, somewhat above 20 p.d.u. For Po-, Pb- and Hg- isotopes below N=126, we observe
lower values of p.d.u. somewhat above 10 p.d.u. For the case of the Po isotopes, there are two branches 
below $N=126$, where one branch is hindered in the decay due to configuration changes, as shown
in Ref. \cite{Kar06,Delion14}.
The reduced width of Pb and Hg isotopes reflects the restricted configuration space for cluster formation,  particular due to 
the protons being at or just below shell closure. For $N>126$, and $Z>82$, neutrons and protons are above the shell gap opening a wide 
configuration space for cluster formation.



We have evaluated the  $\alpha$ decay formation amplitude for the doubly magic nucleus $^{208}$Pb, which is stable due to low $Q$ value,
following the microscopic treatment as described in Ref. \cite{for}. What is striking is that the calculated  $\alpha$ decay 
formation amplitude is nearly unity in p.d.u. This result is quite reasonable since one expects minimal collectivity in  
the nucleus $^{208}$Pb. It further validates the approximation we applied in deriving Eq. (\ref{pdu}). 
A strong reduction can also be expected for the decays from non-collective high-spin isomeric states \cite{Car14}. 

\begin{figure}
		\begin{center}
		\includegraphics[width=0.45\textwidth]{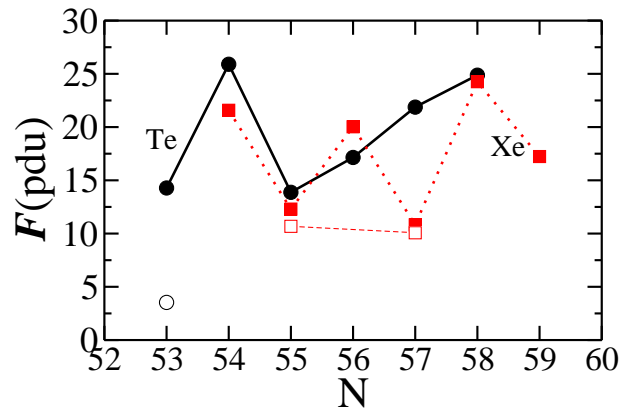}
	\end{center}
	\vspace{-0.6cm}
	\caption{(Color online) $\alpha$-decay formation amplitude in p.d.u as a function $N$ for neutron-deficient 
		Te (circle) and Xe (square) above $^{100}$Sn. Open symbols correspond to the decays 
		of $\alpha$ particles carrying orbital angular momentum $l=2$.  The experimental data  are extracted 
from Ref. \cite{nudat,PhysRevLett.105.162502,Auranen2018}.
		\label{fvsrpnz}}
\end{figure}

In Fig. \ref{fvsrpnz} we compare the p.d.u. of $\alpha$ formation amplitudes of nuclei 
above $^{100}$Sn. The decay widths of those nuclei have attracted significant attention in recent studies in 
relation to the search for the so-called superallowed  $\alpha$
decay. This is expected due to the enhanced neutron-proton pairing when approaching 
the $N=Z$ line and hence an enhanced clustering effect \cite{Lid06,Sew06,Auranen2018,Clark}. One can see from Fig. 
\ref{fvsrpnz} that the formation amplitude of 
those nuclei follows the general average trend of $\alpha$ formation 
amplitude systematics 
even though it shows rather large fluctuations and uncertainties. 
Further experimental investigations
are essential to clarify this issue. It may be useful to mention here that the 
systematics of formation probabilities for available $\alpha$ decays shows an 
increasing trend with decreasing mass number. Apparently, as our formula for p.d.u.
shows, the formation of $\alpha$ scales with the nuclear volume, $~1/A$.
This important feature revealed by our results, needs to be taken into account in studies of $\alpha$ decays 
of trans-tin nuclei, in particular when comparing to heavy nuclei including in particular $^{212}$Po.

\begin{figure}
	\begin{center}
		\setlength{\abovecaptionskip}{0.0cm}
		\includegraphics[width=0.53\textwidth]{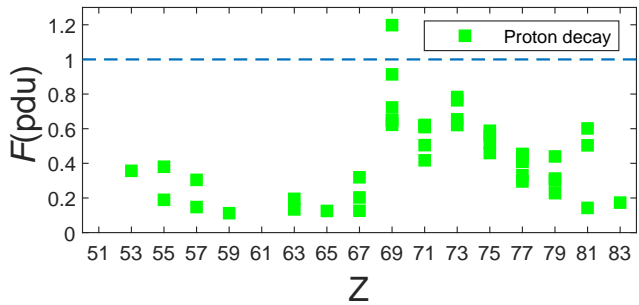}
		\caption{\label{proton} Proton decay formation amplitude in p.d.u. extracted from known data \cite{Delion06,Wang17} on decays from ground states and low-lying isomeric states.}
	\end{center}
\end{figure}

One can employ the $\alpha$ decay formation amplitude in Eq.~(\ref{pdu}) to 
go one step further and evaluate even the $\alpha$ decay width in p.d.u. This is an 
extension of the Weisskopf prescription. In electromagnetic transitions one
evaluates the $B(E\lambda/M\lambda)$ values (which is the equivalent of the alpha formation 
amplitude in our case) from the decay half-life/width by excluding the effect of the decay energy. 
In our case of particle decay, this 
can be easily performed by using Eq.~(\ref{widef}). Thus, the decay width
in p.d.u. is, 
\begin{eqnarray}\label{pdu-w}
\Gamma_{\alpha;{\rm pdu}} (R) &=& \frac{\hbar^2 k}{\mu}
\frac{R^2{ F}^2_{\alpha;{\rm pdu}}(R)}{|H_l^+(\chi,\rho)|^2}\nonumber\\
&\approx& \frac{\hbar^2 k}{\mu}
\frac{R^2{ F}^2_{\alpha;{\rm pdu}}(R)}{{ {\Huge e }^{2\left[{\pi}/{2}-
		2\left({\rho}/{\chi}\right)^{1/2} + 1/3
		\left({\rho}/{\chi}\right)^{3/2} \right]\cdot\cot\beta}}}\end{eqnarray}
where $\cos^2\beta={\rho}/{\chi}$. The derived width above  depends now upon the decay $Q$-value.
For details of the approximate form of the Coulomb 
function used in this equation, where $l=0$ was assumed, see Ref. \cite{udl}.

Our derivation can also be extended to other decay processes including heavier cluster decays and decays that involve change of 
angular momentum. Similarly, we can evaluate  
proton decay, where the formation amplitude becomes much simpler 
than for  $\alpha$ decay in Eq. (\ref{foram}) 
since it involves no intrinsic structure (for details, see Ref. \cite{clw}). With the same assumption as above for the single-particle 
wave function, the proton decay formation amplitude results to have the
simple form of
\begin{equation}\label{pformation}
{ F}_{p;{\rm pdu}} (R)=\frac{1}{R^{3/2}}.
\end{equation}
Using this value, we depict proton emitters in Fig.~\ref{proton}. As expected, since the proton already is 
formed inside the nucleus, the 
p.d.u. corresponding to unity sets the limit for the decay. Values smaller then one indicate a partial occupation
of the particular proton-emitting $l$-state in the daughter nuclei
and/or a change in structure/deformation between mother and daughter nuclei.
Most decays in the figure show p.d.u. values between 0.1 and 0.8. The largest two values 
correspond to the proton decays from the odd-odd nuclei $^{144,146}$Tm which
is enhanced due to
the coupling of the decaying proton with the odd neutron \cite{Ferr01}.

In conclusion, we have deduced a simple averaged single particle limit for the $\alpha$ decay formation amplitude and 
decay width, which we call particle decay unit (p.d.u.). This definition enables a unified description and comparison 
of $\alpha$ decay along the nuclear chart. The magnitude of the p.d.u.  reveals the collectivity of  $\alpha$ decay. 
The decay pattern nicely reveals that a truly microscopic 
description requires the explicit presence of  $\alpha$ clustering elements at the nuclear surface. 
An important feature revealed by our formalism is that the $\alpha$ formation amplitude in p.d.u. scales with the nuclear volume.
Competing decay mechanisms
within the same mother nucleus can be understood as changes of  $\alpha$ collectivity at the surface.  One may expect a similar 
effect as induced by the competition between pairing and deformation 
in two-nucleon transfer reactions (see, e.g., Ref. \cite{Lay19}). 
Our derivation can be extended to other decay processes including proton decay.
We also hope the definition presented in this paper can be useful for quantifying the role played by  
$\alpha$ clustering in heavy nuclei, which may be expected to exhibit a strong correlation to the slope of the nuclear 
symmetry energy and the underlying nuclear equation of state \cite{Typel}. We presume that
the present definition will greatly enhance the understanding of  $\alpha$ correlations as a probe to nuclear interaction.


\begin{thebibliography}{90}
	
	\bibitem{gam28}
	G. Gamow, Z. Phys. {\bf 51}, 204 (1928).
		\bibitem{Lovas} R. G. Lovas, R. J. Liotta, A. Insolia, K. Varga, and D.
	S. Delion, Phys. Rep. 294, 265 (1998).
	\bibitem{clw}
	C. Qi, R.J.Liotta and R. Wyss, Prog. Part. Nucl. Phys. 
	{\bf 105}, 214 (2019).
\bibitem{VDA}
P. Van Duppen, A. N. Andreyev, The Euroschool on Exotic Beams 5,  65, 2018
https://www.springer.com/gp/book/9783319748771



		\bibitem{wei}
	J. M. Blatt and V. F. Weisskopf, Theoretical Nuclear Physics, John Wiley and
	Sons, New York, p. 623  (1952).
	\bibitem{Ras59}] J. O. Rasmussen, Phys. Rev. 113, 1593 (1959).
	\bibitem{Hil19}J. Hilton et al.
	Phys. Rev. C 100, 014305 (2019).
	\bibitem{Cub20}J. G. Cubiss et al.
	Phys. Rev. C 101, 014314 (2020). 

	\bibitem{Thom}R. G. Thomas, Prog. Theor. Phys. 12, 253 (1954).
	\bibitem{Mang}H. J. Mang, Phys. Rev. 119, 1069 (1960).
 \bibitem{TonAr}I. Tonozuka and A. Arima, Nucl. Phys. A323 45 (1979).
 \bibitem{Delion92}D. S. Delion, A. Insolia, and R. J. Liotta
 Phys. Rev. C 46, 1346 (1992).

	\bibitem{lev}
I. I. Levintov, Nucl. Phys. {\bf 4}, 330 (1957).
\bibitem{for}C. Qi, A. N. Andreyev, M. Huyse, R. J. Liotta, P. Van Duppen, and R. A. Wyss
Phys. Rev. C 81, 064319 (2010).
\bibitem{And13}A. N. Andreyev et al., Phys. Rev. Lett. 110, 242502 (2013).
		\bibitem{dls}
D. S. Delion et. al. Phys. Rev. C {\bf 85}, 064306 (2012).
	\bibitem{Kar06}D. Karlgren, R. J. Liotta, R. Wyss, M. Huyse, K. Van de Vel, and P. Van Duppen
	Phys. Rev. C 73, 064304 (2006).
\bibitem{Delion14}D. S. Delion, R. J. Liotta, C. Qi, and R. Wyss
Phys. Rev. C 90, 061303(R) (2014).

\bibitem{Car14}R.J. Carroll et al., Phys. Rev. Lett. 112, 092501 (2014).
	\bibitem{Lid06}S. N. Liddick et al.,
	Phys. Rev. Lett. 97, 082501 (2006).
\bibitem{Sew06}D. Seweryniak et al.,
Phys. Rev. C 73, 061301(R) (2006)
	\bibitem{PhysRevLett.105.162502} I. G. Darby et al.,
	Phys. Rev. Lett. 105, 162502 (2010).
	\bibitem{Auranen2018} K. Auranen et al.,
	Phys. Rev. Lett. 121, 182501 (2018).
		\bibitem{nudat} URL http://www.nndc.bnl.gov/nudat2/
		\bibitem{Clark}R. Clark, A. O. Macchiavelli, H. L. Crawford, P. Fallon, D. Rudolph, A. Såmark-Roth, C. M. Campbell, M. Cromaz, C. Morse, and C. Santamaria, to appear in Phys. Rev. C
		
		\bibitem{Delion06} D.S. Delion, R.J. Liotta, and R. Wyss, Phys. Rep. 424,
		113 (2006).
		\bibitem{Wang17}F. Wang et al, Phys. Lett. B 770, 83 (2017).
		\bibitem{Ferr01}L.S. Ferreira and E. Maglione,
		Phys. Rev. Lett. 86, 1721 (2001).
		\bibitem{udl}C. Qi, F. R. Xu, R. J. Liotta, R. Wyss, M. Y. Zhang, C. Asawatangtrakuldee, and D. Hu
		Phys. Rev. C 80, 044326 (2009);  C. Qi, F.R. Xu, R.J. Liotta, and R. Wyss, Phys. Rev. Lett. 103, 072501 (2009).
		
		\bibitem{Lay19}J.A. Lay, A. Vitturi, L. Fortunato, Y. Tsunoda, T. Togashi, T. Otsuka, 	arXiv:1905.12976.
			\bibitem{Typel}S. Typel,
		Phys. Rev. C 89, 064321 (2014).
\end{thebibliography}
\end{document}